\documentclass[aps,prl,twocolumn,superscriptaddress,floatfix]{revtex4-2}
\usepackage{amsmath,amssymb,graphicx}
\graphicspath{{../../figures/}{../figures/}}
\newcommand{\avg}[1]{\langle #1\rangle}
\newcommand{\D}{\mathcal{D}}
\begin{document}

\title{Synchrony by Birth and Death}

\author{K. P. O'Keeffe}
\email{kevin.p.okeeffe@gmail.com}
\affiliation{Starling Research Institute, Seattle, WA 98112, USA}

\begin{abstract}
A population of oscillators typically synchronizes because coupling pulls their phases together.
Here we consider malthusian oscillators whose coupling is demographic: oscillators are born and
die at rates determined by their phases, generating an effective coupling without any phase
velocity interaction. We find this coupling can synchronize a population, select a collective frequency, and produce a nongeneric fourth-root onset of coherence.
These collective dynamics admit an exact reduction: a population with $m$ frequency classes reduces to $3m$ ordinary differential equations. This is the malthusian analogue of the Ott--Antonsen
reduction for Kuramoto oscillators.
\end{abstract}
\maketitle

Coupled oscillators normally synchronize by phase pulling: coupling advances the slow ones and
retards the fast ones until they
lock~\cite{winfree1967,kuramoto1984,strogatz2000,pikovsky2001,acebron2005}.  But sometimes coupling acts through
growth rates instead.  In cyanobacteria, circadian phase determines when cells are allowed to divide~\cite{mori1996}.
In mammalian cells, cell-cycle phase biases which divide and which die~\cite{chakrabarti2018}.
In yeast, metabolic phase controls when replication can occur~\cite{tu2005}.  In each case the oscillator's phase controls its survival and reproduction, not its velocity.
Prior work coupled growth to phase, finding it disrupts synchrony~\cite{wood2006,yu2015}; here there is no phase coupling to disrupt.
Can a population synchronize when coupling acts only through birth and death?

Here we show it can.  We call such units malthusian oscillators: their coupling acts through per-capita growth rates, not phase velocities.  Their dynamics admit an exact reduction analogous to Ott--Antonsen (OA) theory~\cite{ott2008,ott2009}: just as sinusoidal phase coupling preserves Poisson kernels in OA, sinusoidal birth and death rates preserve von Mises densities, collapsing the infinite-dimensional population dynamics to a finite ODE system on a globally attracting manifold.  The consequences are striking.  Birth and death alone can generate synchrony.  When oscillators carry different inherited frequencies, demographic selection concentrates the population near one rhythm---and cancels the usual cubic saturation, so coherence grows as a fourth root of the coupling rather than the usual square root.

\begin{figure}[t]
\centering
\includegraphics[width=\columnwidth]{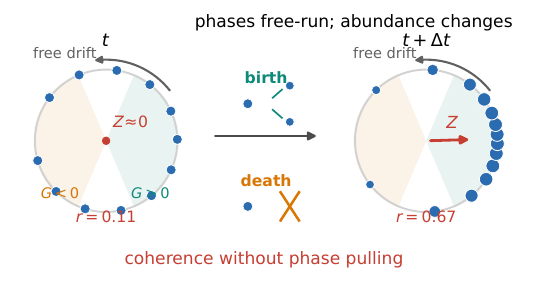}
\caption{Synchrony by birth and death.  Starting from a near-uniform population (left, $|Z|\approx 0$), oscillators aligned with the emerging mean field divide while opposed ones die.}
\label{fig:mechanism}
\end{figure}

\emph{Model.}---A malthusian oscillator carries an internal phase $\theta$ advancing at its natural frequency $\dot\theta=\omega$; it can be born and die, with offspring inheriting the parent's phase and frequency.  A population of such oscillators is
described by the density $n(\theta,\omega,t)$, the number of oscillators near phase
$\theta$ with frequency $\omega$ at time $t$.  We consider $m$ discrete frequency classes with
frequencies $\omega_a$, so $n_a(\theta,t)>0$ is the density of class $a$, whose
members drift at $\dot\theta=\omega_a$.  The population obeys
\begin{align}
 \partial_t n_a&=-\omega_a\partial_\theta n_a+G_a n_a,
 \label{eq:abundance}\\
 G_a&=\underbrace{K r\cos(\theta-\Phi)}_{\text{alignment fitness}}
 -\underbrace{\gamma\ln\!\left(\frac{2\pi n_a}{N_{*,a}}\right)}_{\text{crowding penalty}},
 \label{eq:growth}
\end{align}
where $Z=re^{i\Phi}$ is the mean-field order parameter, $N_{*,a}>0$ is a reference abundance,
$K\ge0$, and $\gamma>0$.  Here $G_a$ is the per-capita growth rate.  It has two parts.  The first is an alignment
fitness: an oscillator whose phase points along the mean field $Z$ is more likely to divide
[Fig.~\ref{fig:mechanism}].
The second is a crowding penalty: phases that are locally overrepresented are suppressed,
preventing indefinite concentration.  We choose the Gompertz law $-\gamma\ln(2\pi n_a/N_{*,a})$ as the crowding penalty: it is analytically convenient and other choices give qualitatively the same results (Supplemental Material).
Neither term appears in $\dot\theta$---the coupling reshapes who survives, never how fast anyone spins.

Let $N_a$ be the total abundance of class $a$, $p_a$ its fraction of the population, and
$\rho_a=n_a/N_a$ its normalized phase density.  The mean field $Z$ is the phasor average
weighted by the actual, evolving population fractions:
\begin{align}
 N_a&=\int_0^{2\pi}n_a\,d\theta,
 &p_a&=\frac{N_a}{\sum_bN_b},
 &\rho_a&=\frac{n_a}{N_a},\notag\\
 Z&=\sum_a p_a Z_a,
 &Z_a&=\int_0^{2\pi}e^{i\theta}\rho_a\,d\theta .
 \label{eq:definitions}
\end{align}

To separate phase and abundance dynamics, integrate Eq.~\eqref{eq:abundance} over $\theta$ to
get $\dot N_a=\avg{G_a}_aN_a$.  Differentiating $\rho_a=n_a/N_a$ then gives the governing density equation
\begin{equation}
 \partial_t\rho_a=-\omega_a\partial_\theta\rho_a
 +(G_a-\avg{G_a}_a)\rho_a .
 \label{eq:conditional}
\end{equation}

\emph{Identical frequencies.}---Start with the simplest case: all oscillators share the same
frequency $\omega$.  The growth term in Eq.~\eqref{eq:conditional} is $Kr\cos(\theta-\Phi)$ minus its mean, which reweights $\rho$ by $\exp[Kr\cos(\theta-\Phi)]$, a pure first harmonic in $\theta$.  Reweighting a density by the exponential of a first harmonic is precisely the operation under which the von Mises family---the maximum-entropy density at fixed first circular moment~\cite{mardia1999}---is closed.  This is the dual of the OA
closure: there a sinusoidal \emph{drift} preserves the Poisson kernel, here a sinusoidal
\emph{growth rate} preserves the von Mises density.  The conditional density therefore stays
von Mises for all time,
\begin{equation}
 \rho(\theta,t)=
 \frac{\exp[\kappa(t)\cos(\theta-\mu(t))]}
 {2\pi I_0(\kappa(t))} ,
 \label{eq:vm}
\end{equation}
where $\kappa\ge0$ is the concentration and $\mu$ is the mean phase, so $Z=Re^{i\mu}$ with
$R=I_1(\kappa)/I_0(\kappa)$.  Since there is only one class, $r=R$ and $\Phi=\mu$.  Writing $\ell=\ln(N/N_*)$ and projecting onto the first harmonic in Eqs.~\eqref{eq:abundance}--\eqref{eq:conditional} gives
\begin{align}
 \dot\kappa &= -\gamma\kappa + KR,\qquad R=\frac{I_1(\kappa)}{I_0(\kappa)},
 \label{eq:kappa}\\
 \dot\mu &= \omega,
 \notag\\
 \dot\ell &= KR^2 - \gamma[\ell+\D(\kappa)],\qquad
 \D(q)=q\frac{I_1(q)}{I_0(q)}-\ln I_0(q).
 \notag
\end{align}
The mean phase drifts freely at $\omega$ and $\ell$ relaxes to a function of $\kappa$, so the
dynamics are governed entirely by the single scalar Eq.~\eqref{eq:kappa}.  Incoherence
($\kappa=0$) is the only fixed point for $K<K_c=2\gamma$; above $K_c$ a synchronized von Mises
state bifurcates supercritically.  Its exact steady state is given implicitly by
\begin{equation}
 R = \frac{I_1(\kappa)}{I_0(\kappa)}, \qquad \kappa = \frac{K}{\gamma}R,
 \label{eq:steady}
\end{equation}
with $R\sim[2(K-K_c)/K_c]^{1/2}$ near onset.

\begin{figure}[t]
\centering
\includegraphics[width=\columnwidth]{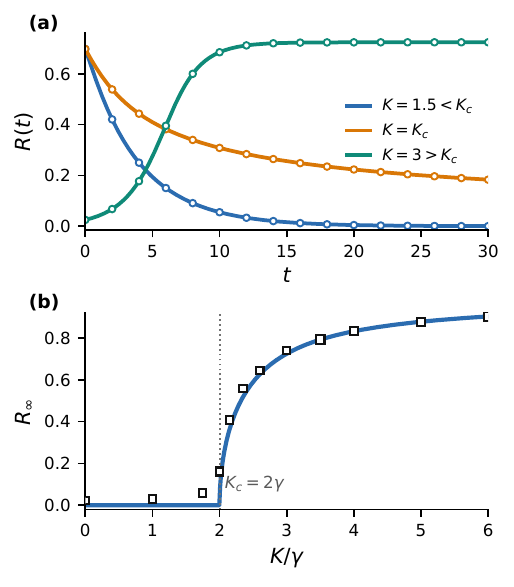}
\caption{Identical malthusian oscillators ($\omega=0$, $\gamma=1$).  (a) Below onset $R(t)$
decays exponentially, at onset it decays algebraically, and above onset it approaches a coherent
state.  Curves show Eq.~\eqref{eq:kappa}; symbols show the full kinetic equation.  (b) The exact
stationary branch (curve) and stochastic birth--death simulation ($N\approx10^3$, symbols).}
\label{fig:identical}
\end{figure}

Figure~\ref{fig:identical} shows all three dynamical regimes and the steady branch.  At criticality,
Eq.~\eqref{eq:kappa} gives $\dot\kappa\sim-\gamma\kappa^3/8$, hence
$R(t)\sim(\gamma t)^{-1/2}$.
Equations~\eqref{eq:kappa}--\eqref{eq:steady} are confirmed in Fig.~\ref{fig:identical}.

\emph{Non-identical oscillators.}---The closure extends immediately to $m$ frequency classes:
each $\rho_a$ stays von Mises with concentration $\kappa_a$, mean phase $\mu_a$, and order
parameter $Z_a=R_ae^{i\mu_a}$.  Writing
$\ell_a=\ln(N_a/N_{*,a})$ gives the exact closed flow
\begin{align}
 \dot\kappa_a &= -\gamma\kappa_a + Kr\cos(\mu_a-\Phi),
 \label{eq:kappaflow}\\
 \dot\mu_a &= \omega_a - \frac{Kr\sin(\mu_a-\Phi)}{\kappa_a},
 \label{eq:muflow}\\
 \dot\ell_a &= KrR_a\cos(\mu_a-\Phi)
 -\gamma[\ell_a+\D(\kappa_a)],
 \label{eq:ellflow}
\end{align}
where $Z=re^{i\Phi}=\sum_a p_a R_a e^{i\mu_a}$.  So $m$ classes reduce to $3m$ real
variables, coupled only through the mean field.  Despite appearances, Eq.~\eqref{eq:muflow} is not phase pulling: no oscillator deviates from
$\dot\theta=\omega_a$; what moves is the class mean, shifted by birth and death reweighting the
density.  The von Mises manifold is globally attracting among smooth positive densities: every transverse
mode decays at rate $\gamma$, independent of coupling and abundances (Supplemental Material).

\emph{Two frequencies.}---We now use Eqs.~\eqref{eq:kappaflow}--\eqref{eq:ellflow} to study
two classes at $\omega_1=+\omega_0$ and $\omega_2=-\omega_0$, with equal initial abundances.
We work in the frame following the mean phase $\Phi$, so $\Omega=\dot\Phi$ enforces $Z=r>0$.
The real variables are $(\kappa_1,\kappa_2,\mu_1,\mu_2,\delta)$, where the $\mu_j$ are measured
from $\Phi$ and $\delta=\ell_1-\ell_2$:
\begin{align}
 \dot\kappa_i &= -\gamma\kappa_i + Kr\cos\mu_i, \quad i=1,2
 \label{eq:kappasys}\\
 \dot\mu_i &= \omega_i-\Omega - \frac{Kr\sin\mu_i}{\kappa_i}, \quad i=1,2
 \label{eq:musys}\\
 \dot\delta &= Kr[R_1\cos\mu_1-R_2\cos\mu_2]
 -\gamma[\delta+\D(\kappa_1)-\D(\kappa_2)],
 \label{eq:deltasys}
\end{align}
where $R_j=I_1(\kappa_j)/I_0(\kappa_j)$, $p_1=e^\delta/(1+e^\delta)$, and $p_2=1-p_1$.  The
mean field $r$ obeys
\begin{align}
 r &= p_1 R_1\cos\mu_1 + p_2 R_2\cos\mu_2,\notag\\
 0 &= p_1 R_1\sin\mu_1 + p_2 R_2\sin\mu_2.
 \label{eq:rself}
\end{align}
Differentiating the second condition determines $\Omega$.  The system admits three steady states.

First is incoherence: $r=0$, $\kappa_1=\kappa_2=0$, $\delta=0$.  Abundance dynamics decouple at linear
order.  Linearization yields the eigenvalues
\begin{equation}
 \lambda_\pm=-\gamma+\frac K4
 \pm\sqrt{\left(\frac K4\right)^2-\omega_0^2}.
 \label{eq:eigs}
\end{equation}
Thus the critcal coupling for incoherence iss
\begin{equation}
 K_{\rm inc}=\begin{cases}
 K_{\rm s}=2(\gamma^2+\omega_0^2)/\gamma,&\omega_0<\gamma,\\
 K_{\rm o}=4\gamma,&\omega_0>\gamma.
 \end{cases}
 \label{eq:Ks}
\end{equation}
For $\omega_0<\gamma$, a real eigenvalue vanishes at $K_{\rm s}$ and incoherence undergoes a
supercritical steady bifurcation.  For $\omega_0>\gamma$, two symmetry-related complex pairs cross
at $K_{\rm o}$ with frequency $(\omega_0^2-\gamma^2)^{1/2}$, giving an $O(2)$-Hopf bifurcation.

\begin{figure}[t]
\centering
\includegraphics[width=\columnwidth]{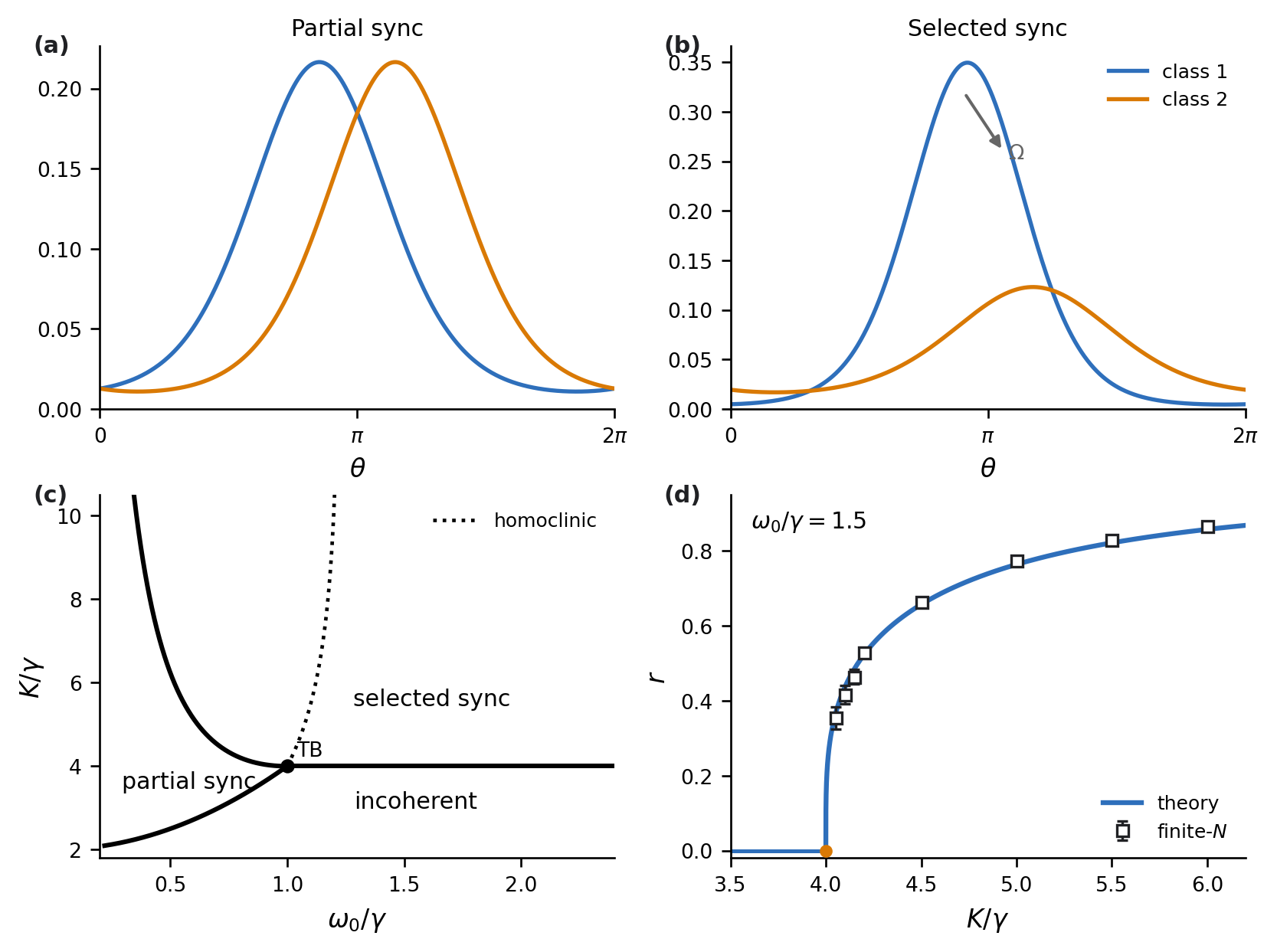}
\caption{Two equal initial populations.
(a) Partial sync: two von Mises densities symmetrically offset.
(b) Selected sync: one class dominates and the population drifts at $\Omega$.
(c) Bifurcation diagram.
(d) Fourth-root onset confirmed by stochastic simulation ($N\approx3$--$4\times10^4$).}
\label{fig:phase_diagrams}
\end{figure}

Second is the partial sync state: $\kappa_1=\kappa_2$, $\mu_1=-\mu_2$, $\delta=0$,
born at $K_{\rm s}$ for $\omega_0<\gamma$ [Fig.~\ref{fig:phase_diagrams}(a)].  Let $\nu_j=\omega_j-\Omega$.  Then
Eqs.~\eqref{eq:kappasys}--\eqref{eq:musys} give
\begin{equation}
 \kappa_j=\frac{Kr}{d_j},\qquad
 \tan\mu_j=\frac{\nu_j}{\gamma},\qquad
 d_j=(\gamma^2+\nu_j^2)^{1/2},
 \label{eq:relative}
\end{equation}
and Eq.~\eqref{eq:deltasys} reduces to
\begin{equation}
 p_j=\frac{N_{*,j}I_0(\kappa_j)}{\sum_k N_{*,k}I_0(\kappa_k)}.
 \label{eq:selectionweight}
\end{equation}
A class closer to the collective rhythm has larger $\kappa_j$ and is therefore more populous.  One eigenvalue vanishes when
$I_1(\kappa)/[\kappa I_0(\kappa)]=\omega_0^2/(\gamma^2+\omega_0^2)$, giving
\begin{equation}
 K_{\rm sel}=
 \frac{(\gamma^2+\omega_0^2)^2}{\gamma\omega_0^2},
 \qquad 0<\omega_0<\gamma,
 \label{eq:Ksel}
\end{equation}
via a supercritical drift pitchfork: the partial sync state loses reflection symmetry and splits
into two counter-rotating states [Fig.~\ref{fig:phase_diagrams}(c)].

Third is the selected sync state: $\delta\ne0$ and $\Omega\ne0$, born at
$K_{\rm sel}$ [Fig.~\ref{fig:phase_diagrams}(b)].
One class takes over and the population rotates at its inherited frequency.  For $\omega_0>\gamma$ the transition is instead an $O(2)$-Hopf at $K_{\rm o}=4\gamma$, where two counter-rotating selected states and an unstable standing wave are born together.  On the selected branch, demographic reweighting cancels the cubic saturation, so coherence grows as a fourth root of the coupling.  With $\omega_0/\gamma>1$,
\begin{align}
 r^4&=\frac{3(\omega_0/\gamma)^2}{4(2(\omega_0/\gamma)^2+1)}
 \left(\frac K\gamma-4\right)
 +O[(K/\gamma-4)^{3/2}],
 \label{eq:fourthroot}
\end{align}
For each fixed $\omega_0/\gamma>1$ the prefactor is positive, so the selected branch has fourth-root onset; the
expansion is nonuniform as $\omega_0/\gamma\downarrow1$.  Exact continuation, direct integration, and a stochastic
simulation confirm the exact branch [Fig.~\ref{fig:phase_diagrams}(d)].  The three
boundaries meet at $(\omega_0,K)=(\gamma,4\gamma)$, a Takens--Bogdanov point with $O(2)$ symmetry~\cite{dangelmayr1987}.  This corner is degenerate: the same demographic cancellation that produces the fourth root also annihilates the cubic coefficient of the traveling wave along the entire line $K=4\gamma$, $\omega_0>\gamma$.  As a result the generic Takens--Bogdanov unfolding does not apply, and a curve of Bautin degeneracies~\cite{kuznetsov2004} runs into the corner.  Inside the reflection-symmetric subspace the standing wave is destroyed by a gluing bifurcation followed by a saddle-node of periodic orbits.

\begin{figure}[t]
\centering
\includegraphics[width=\columnwidth]{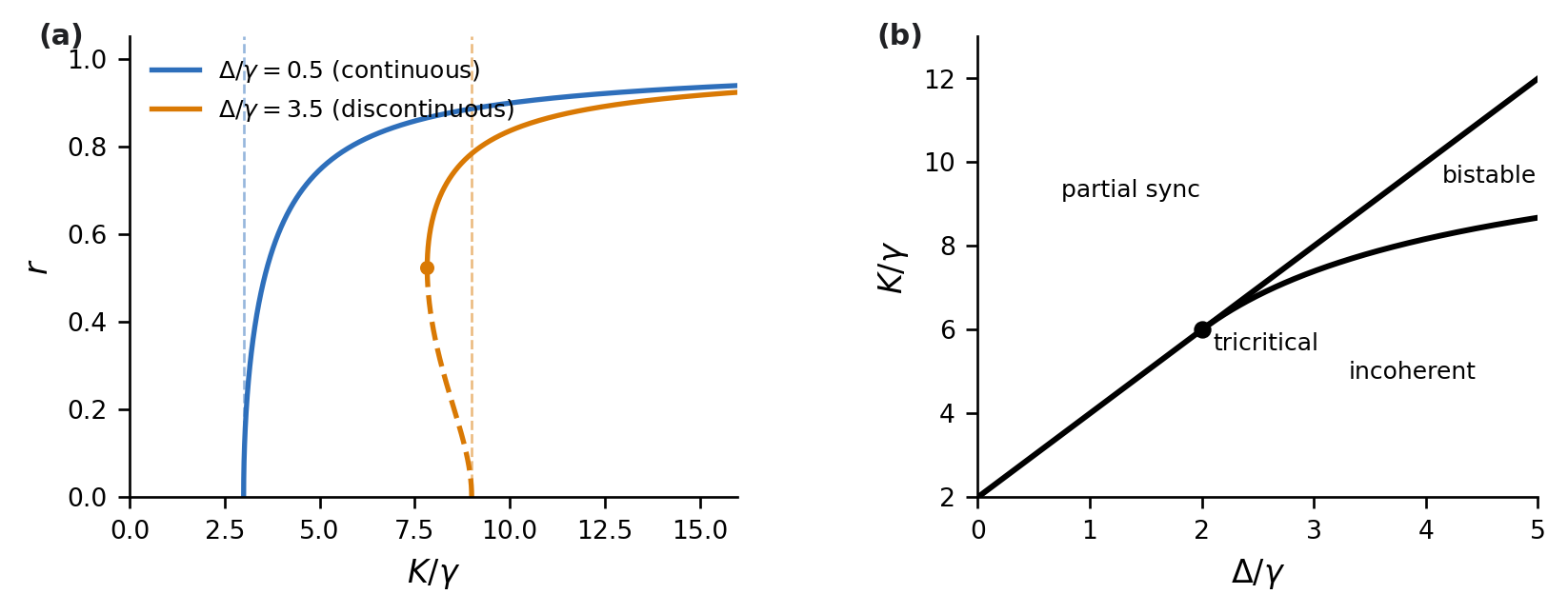}
\caption{Lorentzian frequency continuum.  (a) Order parameter $r$ vs $K/\gamma$ for a
continuous ($\Delta/\gamma=0.5$) and discontinuous ($\Delta/\gamma=3.5$) transition; solid:
stable, dashed: unstable, dot: saddle node $K_{\rm sn}$. For $K_{\rm sn}<K<K_c$ the coherent
branch coexists with incoherence.  (b) Onset $K_c$ (upper) and saddle node $K_{\rm sn}$ (lower)
vs $\Delta/\gamma$; both states stable between them.  Curves meet at the tricritical point
$\Delta=2\gamma$.}
\label{fig:branch}
\end{figure}

\emph{Continuum frequencies.}---The von Mises closure extends to a continuous frequency distribution.  Replace
the discrete classes by a Lorentzian
$g_\Delta(\omega)=\Delta/[\pi(\omega^2+\Delta^2)]$.  At steady state each class sits on its von
Mises fixed point with $\kappa_\omega=Kr/d_\omega$; integrating over $g_\Delta$ gives a single
self-consistency equation that fixes the order parameter of every stationary coherent state,
\begin{equation}
 r=\gamma\,
 \frac{\displaystyle\int_{-\infty}^{\infty}
 g_\Delta(\omega)\,\frac{I_1(Kr/d_\omega)}{d_\omega}\,d\omega}
 {\displaystyle\int_{-\infty}^{\infty}
 g_\Delta(\omega)\,I_0(Kr/d_\omega)d\omega},
 \qquad d_\omega=(\gamma^2+\omega^2)^{1/2}.
 \label{eq:continuous_branch}
\end{equation}
Thus Eq.~\eqref{eq:continuous_branch} predicts the full curve $r(K)$, not just its onset
[Fig.~\ref{fig:branch}(a)].
The denominator is the demographic reweighting factor: classes near $\omega=0$ become more
populous, narrowing the active frequency distribution around its one available rhythm.
Linearization about incoherence gives $\lambda=K/2-(\gamma+\Delta)$, hence
\begin{equation}
 K_c=2(\gamma+\Delta).
 \label{eq:continuous_onset}
\end{equation}
The transition has three regimes depending on $\Delta/\gamma$.  For $\Delta<2\gamma$ it is supercritical,
\begin{equation}
 r^2\sim\frac{4\gamma^2}{(\gamma+\Delta)(2\gamma-\Delta)}
 \left(\frac K{K_c}-1\right).
 \label{eq:lorentzian_onset}
\end{equation}
At $\Delta=2\gamma$ demographic reweighting exactly cancels the cubic saturation, giving a fourth-root onset $r^4\sim(16/15)(K/K_c-1)$ at $K_c=6\gamma$ --- the same mechanism as on the selected sync branch.  For $\Delta>2\gamma$ the bifurcation is subcritical: an unstable branch turns at a saddle node $K_{\rm sn}$, producing bistability with incoherence~\cite{yu2015}.  Figure~\ref{fig:branch}(b) shows the full bifurcation diagram as a function of $\Delta/\gamma$.

The selected sync state does not survive a unimodal continuum: demographic feedback narrows a single peak but cannot select among competing rhythms, so $\Omega=0$ is the only collective frequency.  The Supplemental Material gives results for arbitrary frequency distributions and the stability analysis.

\emph{Discussion.}---Our main result is that birth and death alone can induce different forms of synchrony: a simple coherent state, a selected rhythm, and a tricritical transition between continuous and discontinuous onset.  Moreover, we have presented an exact reduction which allows us to describe these phenomena analytically.  These phenomena may be realized in systems where phase controls survival, such as cyanobacteria, mammalian cell cycles, and yeast metabolism.

These results are robust to changes in the crowding law. In the Supplemental Material we replace the Gompertz penalty with the Box--Cox family $\mathcal{C}_s$, which interpolates between softer ($s<0$) and stiffer ($s>0$) penalties. The onset $K_{\rm inc}$ is unchanged for all $s$, and coherence and rhythm selection persist. What changes is the character of the onset: $s>0$ gives a square-root onset, $s=0$ the fourth root, and $s<0$ a subcritical transition with bistability.

Finally, the von Mises closure connects to earlier work.  Exponential families are the natural invariant objects of selection dynamics~\cite{karev2010}, but in Karev's setting the manifold depends on the initial distribution; here it is forgotten at rate $\gamma$.  Unlike OA, whose manifold requires frequency spread, the malthusian manifold is attracting at rate $\gamma$ for any $K$, $\omega$, or abundance profile; a Lorentzian continuum does not, however, admit OA's residue collapse to a finite ODE (Supplemental Material).


\begin{thebibliography}{99}
\bibitem{winfree1967} A.\ T.\ Winfree,
J.\ Theor.\ Biol.\ \textbf{16}, 15 (1967).
\bibitem{kuramoto1984} Y.\ Kuramoto, \emph{Chemical Oscillations, Waves, and Turbulence}
(Springer, Berlin, 1984).
\bibitem{strogatz2000} S.\ H.\ Strogatz,
Physica D \textbf{143}, 1 (2000).
\bibitem{pikovsky2001} A.\ Pikovsky, M.\ Rosenblum, and J.\ Kurths,
\emph{Synchronization: A Universal Concept in Nonlinear Sciences}
(Cambridge University Press, Cambridge, 2001).
\bibitem{acebron2005} J.\ A.\ Acebr\'on, L.\ L.\ Bonilla, C.\ J.\ P\'erez Vicente,
F.\ Ritort, and R.\ Spigler,
Rev.\ Mod.\ Phys.\ \textbf{77}, 137 (2005).
\bibitem{mori1996} T. Mori, B. Binder, and C. H. Johnson,
Proc. Natl. Acad. Sci. USA \textbf{93}, 10183 (1996).
\bibitem{chakrabarti2018} S. Chakrabarti \emph{et al.},
Nat. Commun. \textbf{9}, 5372 (2018).
\bibitem{tu2005} B. P. Tu, A. Kudlicki, M. Rowicka, and S. L. McKnight,
Science \textbf{310}, 1152 (2005).
\bibitem{wood2006} K.\ Wood, C.\ Van den Broeck, R.\ Kawai, and K.\ Lindenberg,
Phys.\ Rev.\ Lett.\ \textbf{96}, 145701 (2006).
\bibitem{yu2015} W.\ Yu and K.\ B.\ Wood,
Phys.\ Rev.\ E \textbf{91}, 062708 (2015).
\bibitem{ott2008} E.\ Ott and T.\ M.\ Antonsen,
Chaos \textbf{18}, 037113 (2008).
\bibitem{ott2009} E.\ Ott and T.\ M.\ Antonsen,
Chaos \textbf{19}, 023117 (2009).
\bibitem{mardia1999} K.\ V.\ Mardia and P.\ E.\ Jupp, \emph{Directional Statistics}
(Wiley, Chichester, 1999).
\bibitem{dangelmayr1987} G.\ Dangelmayr and E.\ Knobloch,
Philos.\ Trans.\ R.\ Soc.\ London A \textbf{322}, 243 (1987).
\bibitem{kuznetsov2004} Y.\ A.\ Kuznetsov, \emph{Elements of Applied Bifurcation Theory}, 3rd ed.
(Springer, New York, 2004).
\bibitem{karev2010} G.\ P.\ Karev,
Entropy \textbf{12}, 1673 (2010).
\end{thebibliography}
\end{document}